Artificial Intelligence as an Opportunity for the Science of Consciousness:

A Dual-Resolution Framework

Shahar Dror[1], Dafna Bergerbest[2], Moti Salti[1]

[1]Department of Cognitive and Brain Sciences, Ben-Gurion University of the Negev; Beer Sheva, 84105, Israel.

[2]School of Behavioral Sciences, The Academic College of Tel-Aviv Yaffo; Tel-Aviv Yaffo, 61083, Israel.

*Corresponding author email: motisalti@gmail.com

Short title: A Dual-Resolution Framework for Consciousness

Abstract

The encounter of artificial intelligence with consciousness research is often framed as a challenge: could this science determine whether such systems are conscious? We suggest it is equally an opportunity to expand and test the scope of existing theories of consciousness. Current approaches remain polarized. Computational functionalism emphasizes abstract organization, often realized through neural correlates of consciousness, while biological naturalism insists that consciousness is tied to living embodiment. Both positions risk anthropocentrism and limit the possibility of recognizing non-biological forms of subjectivity. To move beyond this impasse, we propose a dual-resolution framework that that specifies candidate ontological and epistemic conditions under which consciousness could emerge. This approach combines the Information Theory of Individuality, which characterizes informational autonomy and self-maintaining individuality, with the Moment-to-Moment framework, which characterizes temporal updating and phenomenological unfolding. This integration reframes consciousness as the possible epistemic expression of informationally individuated systems, while positioning AI as a constructive testbed for evaluating whether such systems generate operational markers of subjectivity. In this way, artificial systems become not merely objects of consciousness attribution, but experimental platforms for testing the generality of consciousness theories across biological and non-biological substrates.

The rapid advancements in machine learning have sparked widespread speculation and debate about the possibility of artificial intelligence (AI) achieving consciousness (Butlin et al., 2025; Chalmers, 2023). Naturally, this posed a challenge to scientists who study consciousness, namely, to determine whether these systems are sentient. Early work by Gamez (2008) identifies four areas of machine consciousness research: (MC1) machines with the external behavior associated with consciousness, (MC2) machines with the cognitive characteristics associated with consciousness, (MC3) machines with an architecture claimed to be a cause or correlate of human consciousness, and (MC4) phenomenally conscious machines. Although this taxonomy is helpful in organizing the field, there is an overlap between these categories. For example, modelling cognitive characteristics (MC2) may be used to produce more realistic conscious behavior (MC1) or may be implemented using an architecture linked to consciousness (MC3), and MC3 can also overlap with MC4 if one assumes that implementing such an architecture could yield phenomenal states.

More recently, Chalmers (2023) surveys evidence for and against consciousness in Large Language Models (LLMs), focusing on what Gamez refers to as MC3. Chalmers utilizes the modern scientific study of consciousness. This line of research, established in Crick and Koch's (1990) seminal paper focuses on the neural correlates of consciousness (NCC). Crick and Koch suggested that identifying the NCC through empirical contrasts between conscious and unconscious states would supply an objective handle that would allow scientific study on the subjective and reveal the nature of consciousness. This framework supplied a plethora of theories and complimentary empirical results indicating the brain's activity associated with consciousness (Friedman et al., 2023). Chalmers highlights recurrent processing and global availability of information as central features plausibly required for consciousness. He argues that present-day systems likely lack these features in a sufficient form but maintains that there is no principled barrier to their realization in artificial systems in the future. Butlin et al. (2025) go further addressing this challenge of identifying consciousness in artificial systems through a meticulous and systematic, theory-driven examination. They extracted indicators from leading theories of consciousness, including Global Neuronal Workspace Theory (GNWT),

Integrated Information Theory (IIT), and recurrent processing-based accounts and tested whether these indicators are instantiated by AI systems. Butlin et al. handle indicators in a Bayesian manner, so that the existence of each indicator shifts, rather than settles, the credence that a system is conscious. Respectively, they do not conclude whether current AI systems are conscious; instead, they treat the presence of those indicators in future models as an open research question.

**The lack of a general definition to consciousness**

The question of whether artificial systems can be conscious can be understood as a special case of a broader scientific problem: how to generalize consciousness beyond the human case. This problem is not new. In comparative animal research, studies have already been asking whether, and on what grounds, consciousness can be attributed to non-human systems (Boly et al., 2013). Historically, this work has proceeded along two main paths. One path treats certain non-human animals, especially non-human primates, as model systems for studying consciousness under the assumption that they are conscious. The other asks whether a particular species has conscious experience at all, with the aim of identifying the evolutionary origins and comparative physiological bases of consciousness (Edelman & Seth, 2009; Feinberg & Mallatt, 2013; Mashour & Alkire, 2013).

This second path is, in effect, an attempt to validate consciousness through generalization, but it also reveals the limits of prevailing approaches. Generalization is relatively tractable when species share homologous neural structures, as in many mammals, yet it becomes much harder when candidate systems differ substantially from humans in their physical organization. Across distantly related animals, brain structures and neural circuitry often lack close homology to the human brain, making mechanism-based extrapolation increasingly uncertain. The same difficulty appears even more sharply in the case of artificial systems. If dominant theories operationalize consciousness primarily in terms of human neural architectures and human-specific mechanisms, then their ability to generalize will remain fundamentally constrained.

This points to a deeper conceptual problem: contemporary frameworks of consciousness are often implicitly anthropocentric, taking the human case as primary and treating other instances as derivative. The difficulty is especially visible in the NCC framework that has dominated modern consciousness science since Crick and Koch (1990). The NCC approach aimed to identify the neural conditions that systematically distinguish conscious from unconscious states, but it did so without first settling what consciousness is. Rather than beginning with a priori definition of the phenomenon, the framework proceeded by contrasting cases that were already treated as conscious or unconscious and then searching for their neural correlates. This strategy enabled productive empirical work, but it also left the central concept underdetermined: what exactly counts as consciousness, what falls outside it, and which of its features are essential rather than contingent?

Crick and Koch deliberately left this question open, proposing instead a reciprocal relation between empirical findings and theory, in which discoveries about NCCs would gradually refine the concept of consciousness itself, and revised concepts would in turn generate new empirical predictions. In principle, this iterative process was meant to culminate in a lucid and general account of consciousness (Revach & Salti, 2021). Yet the challenge of extending consciousness beyond humans suggests that this goal has not been fully achieved. As long as consciousness is operationalized through paradigmatic human cases and human neural organization, generalization remains constrained by the very assumptions built into the framework.

From this perspective, artificial consciousness is not a wholly separate problem, but a stringent test of whether consciousness science has developed a genuinely general concept of its object. AI forces the field to confront the unresolved tension at the heart of the NCC program: can consciousness be identified in a way that is not tied in advance to specifically human or even specifically biological mechanisms? If not, then the limits of current theories are not merely empirical but conceptual.

**Does AI present a valid challenge to the scientific study of consciousness?**

Not all would agree that the question of identifying consciousness in artificial systems is a valid one to ask, arguing that consciousness is in principle restricted to biological systems. One line of argument challenges the very intuition that artificial systems might be conscious. Seth (2024) suggests that attributions of consciousness to AI may stem from two well-known psychological biases: anthropocentrism and anthropomorphism. Anthropocentrism, human exceptionalism, leads us to treat distinctively human capacities, especially language and advanced cognition, as benchmarks for consciousness, while anthropomorphism leads us to project human-like mental properties onto non-human systems. Together, these biases predispose us to over-ascribe consciousness to systems that exhibit surface-level human-like behaviors, such as fluent language use or adaptive responses.

Seth (2024) further argues that many discussions of AI consciousness implicitly rely on the philosophical position of computational functionalism (Putnam, 1967; Fodor, 1968; Chalmers, 2011). According to this view, the right kind of computational or functional organization is sufficient for consciousness, independently of the physical substrate in which it is implemented. On this view, if an artificial system implemented the relevant computational organization, its non-biological substrate would not by itself matter. Seth challenges this assumption by questioning whether the functions relevant to consciousness can really be separated from the biological processes that realize them.

Instead of computational functionalism, Seth subscribes to a form of biological naturalism as portrayed by Searle (1992). According to Searle, consciousness arises from the unique biological, biochemical, and embodied processes of living systems. On this view, without these processes, artificial systems may simulate thought but lack the subjective awareness that gives rise to what biological naturalists regard as an authentic being. Accordingly, the development of consciousness is tightly interweaved with the environmental stressors that constrain the living entity. This does not amount to simple carbon chauvinism, nor does it show that conscious artefacts are impossible in principle. Rather, it suggests that real artificial consciousness would require systems that are, in some relevant sense, life-like or biologically organized, rather than merely systems that simulate intelligent behaviour through conventional digital computation. From this perspective, current AI does not present a genuine challenge to the scientific study of consciousness, since they lack the self-maintaining, embodied, and metabolically grounded organization that biological naturalists take to be central to conscious experience.

Another argument is proposed by Buonomano (2026), who emphasizes the fundamentally temporal nature of consciousness. According to Buonomano, the brain is not merely a computational system, but a continuous-time dynamical process whose activity unfolds in real time. He argues that consciousness, like life itself, is a process unfolding in continuous time. From this perspective, digital computers differ fundamentally from biological systems because their operations occur in discrete simulated time and can be paused without affecting the computation. Buonomano therefore suggests that such systems cannot support the continuous temporal dynamics required for conscious experience, particularly the subjective sense of temporal flow. As such, he argues that current digital AI systems cannot genuinely be conscious, while leaving open the possibility that alternative artificial architectures based on continuous-time dynamics may eventually support consciousness. While this account differs from biological naturalism by not tying consciousness directly to biological matter, it still ties consciousness to properties derived from the dynamics of biological systems.

A more moderate position is offered by Milinkovic and Aru (2026), who propose a framework they term biological computationalism. They argue that conscious processing depends on specific computational properties characteristic of biological systems. They identify three central features. The first is hybrid computation, where continuous dynamics and discrete events influence each other. The second is scale-inseparable organization, where processes are integrated across multiple levels, from molecular and cellular activity to population and whole-system dynamics, under metabolic constraints. The third is dynamico-structural co-determination, where the system's functional organization and physical structure continuously shape each other over time.

These features are presented as central to the kind of computation realized in biological systems, and thus as potential prerequisites for consciousness. At the same time, the authors do not treat them as exclusive to biology. They leave open the possibility that artificial systems could, in principle, instantiate similar properties.

Their objection is therefore directed less at artificial consciousness as such than at current digital AI systems, whose computations remain scale-separable, discretized, and largely decoupled from the physical substrate that implements them. However, the connection between these biological-computational features and consciousness itself remains somewhat implicit. Milinković and Aru motivate the connection by appealing to features of conscious experience such as unity, differentiation, temporal coherence, and context-sensitivity, but they do not yet fully establish why hybrid dynamics, scale inseparability, and metabolic embedding should be constitutive of consciousness rather than contingent features of its biological implementation.

Taken together, the positions discussed in this section suggest that the debate over artificial consciousness remains deeply grounded in the properties of biological systems as we currently understand them. Whether by ruling out artificial consciousness in principle, as in biological naturalism, or by identifying specific computational features derived from biological organization, these accounts tie the possibility of consciousness to characteristics observed in known living systems.

As a result, even approaches that leave room for artificial consciousness do so only under conditions modeled on familiar biological processes. Consequently, current digital AI systems, which remain largely insulated from the material and biological pressures that govern living systems, may simulate aspects of cognition without instantiating the conditions necessary for consciousness itself.

**Redefining the relation between consciousness and life**

Biological naturalism proponents claim that consciousness is interweaved with life. Consciousness is either relied on the self-maintaining organization of biological systems (Seth) or by being subjected to the continuous real-time dynamics characteristic of biological systems (Buonomano), or by depending on a specifically biological mode of computation relying on the characteristics of biological substrates (Milinković and Aru). However, could we use a different definition of life that would enable the discussion instead of limiting it? Krakauer and colleagues present the Information Theory of Individuality (ITI), offering an information-theoretic reconstruction of biological individuality, the unit presupposed by life, adaptation, and selection (Krakauer et al., 2020). On their account, individuality is not defined by membranes, replication, or biochemical composition, but by the propagation of information from past to future and the preservation of temporal integrity. This does not reduce life to mere information processing; rather, it shifts the relevant question from whether a system has a familiar biological substrate to whether it forms an informationally autonomous, self-maintaining unit.

The relevant issue here is not whether life should be reduced to information processing in general, but whether the form of organization that makes a system a living individual must be defined biologically. Biological naturalism typically treats consciousness as dependent on the organization of a living subject, yet the notion of such a subject already presupposes an account of individuality: a distinction between the system and its environment, a capacity to preserve organization across time, and some degree of autonomy from immediate external determination. If these features can be described in informational rather than strictly biochemical terms, then the question of artificial consciousness changes form. It becomes not whether an artificial system is biological in the familiar sense, but whether it can constitute a self-maintaining informational individual.

If consciousness depends on life because it depends on the organization of a self-maintaining individual, then ITI weakens the assumption that this individual must be biologically bounded in the familiar organismic sense. The crucial issue becomes whether artificial systems can instantiate informational individuality, not whether they are made of biological tissue.

**The updating approach to consciousness**

Here we elaborate on a framework to consciousness that was first introduced by Salti et al.'s (2019) Moment to Moment (MTM) framework of consciousness. We will first describe the framework. Then we will point to the resemblance in principle of operation between the MTM and ITI (Krakauer et al., 2020). And finally, we will discuss the implications of this resemblance.

Most contemporary NCC research focuses on identifying discrete neural events localized in time and space (see Revach & Salti, 2021, for a critical perspective). In contrast, the MTM framework conceptualizes consciousness as an ongoing dynamic process (Salti et al., 2019). The MTM suggests that consciousness arises from the continuous updating and recoding of stimuli to fit into a perceptual context. Updating, in this sense, does not refer merely to storing new information, but to the transformation of present input considering an already unfolding perceptual history. This updating process is shaped by three main factors: context, stimulus saliency, and observers' goals. According to this view, a stimulus is experienced as it is being actively integrated and updated within the stream of all other perceived stimuli. This process creates hysteresis, an internal trace shaped by the system's unique history, which in turn influences its present responses.

Hysteresis is necessary but not sufficient for subjectivity. A merely history-sensitive system may adapt, predict, or store system's history without thereby possessing a point of view. For hysteresis to become subjectivity-relevant, it must operate within an individuated system, that is, a system whose states are processed relative to a maintained distinction between itself and its environment. This is where the Information Theory of Individuality becomes relevant. Krakauer and colleagues propose that biological individuality is not fundamentally defined by membranes, replication, or biochemical composition, but by the preservation of temporal integrity and the propagation of information from past to future. Once such an individuated boundary is in place, incoming information is not merely classified as external input, it is integrated as a modification of the system's own ongoing trajectory. Each update occurs against the backdrop of a continuous internal sequence of prior updates, so the system recognizes the new input as its own precisely because it must align and 'fit' within that sequence. In this way, the update is experienced as part of the ongoing informational stream that constitutes its 'self.' As a result, a clear distinction emerges between changes attributed to the system itself and changes attributed to the environment, providing the basis for a subjective perspective. The hysteresis is also never identical across systems, not even in those that begin from a similar starting point. Even for identical systems, from the very first moment, spatial differences give rise to divergent experiential trajectories that continue to branch over time. In this way, each system develops a unique, subjective informational stream. Importantly, other characteristics of conscious experience emerge, such as body-boundaries and volition in the same manner (See Revach & Salti, 2022, for details).

Building on this view, we can revisit Nagel's fundamental question: "What is it like to be a bat?" (Nagel, 1974). MTM allows us to understand how different beings may experience the world in qualitatively distinct ways, shaped by their unique constraints, goals, temporal dynamics, and modes of interaction with the environment. For example, even among biological systems, one can imagine vastly different temporal experiences: the phenomenological world of a slow creature such as a sloth may differ substantially from that of a fast creature such as a peregrine falcon, whose sensorimotor capacities and ecological demands unfold on very different time scales. This suggests that consciousness may be deeply shaped by the specific temporal, sensorimotor, and ecological constraints of the system in which it emerges.

From this perspective, the proposed framework offers a way to reconcile different types of consciousness under a common mechanism. A general process of moment-to-moment updating may give rise to diverse forms of subjective experience, each structured by the particular constraints of the system. Thus, if the phenomenological worlds of a sloth and a falcon may already differ profoundly, the subjective experience of an artificial system, organized by radically different temporal, structural, and environmental constraints, may differ even more. When the constraints of two systems diverge, their phenomenological worlds should diverge accordingly; when two systems share similar temporal and structural parameters, their conscious experiences may be more comparable and perhaps even communicable.

According to the MTM, each individual's phenomenological experience would be grounded in its particular trajectory of information updating. In this sense, the updating mechanism described by MTM corresponds to the notion of life depicted by Krakauer and colleagues (2020). The relationship between the mechanisms described by ITI and MTM can be conceptualized as a dual resolution of individuation. ITI provides an **ontological resolution**, specifying the conditions under which a system qualifies as an individual by maintaining informational autonomy, and the constraints operating on this unit. It does so by asking whether the system preserves temporal integrity, propagates information from past to future, and maintains some degree of autonomy from external determination. The mechanism described by MTM, by contrast, provides an **epistemic resolution**, articulating the conditions under which such an individual *appears to itself* as a continuous subject through an updating process that is bound to the constraints of the unit. It does that by updating and recoding each new input in relation to the system's prior trajectory, current context, and goals. In this framework, ITI secures the informational unit to which experience is attributed, while MTM describes the temporal process through which the world becomes available to that unit. Consciousness is therefore not identified with information processing alone, but with the epistemic unfolding of an individuated system.

The dual-resolution framework is especially compatible with theories that view perception and cognition as dynamic and continuously evolving processes, such as Predictive Coding and Active Inference. However, Predictive Processing and Active Inference primarily explain how a system models and regulates its relation to the world. The present framework begins one step earlier, by asking what makes such a system an individuated locus of updating in the first place. It then asks how this updating becomes organized as a subjective trajectory. In this respect, ITI supplies an ontological condition that is often presupposed in predictive-processing accounts: the identification of a system-environment boundary. The MTM component does not merely claim that consciousness unfolds over time, nor does it primarily explain consciousness in terms of prediction, error minimization, global broadcasting, or integrated causal structure. Rather, it proposes that consciousness arises from the way a system continuously updates and recodes information into its own developing trajectory (Salti et al., 2019; Revach & Salti, 2021). Through this process, each new input is integrated against the background of prior updates, generating a history-bound subjective stream. In this sense, consciousness is not simply a temporal process, but the epistemic unfolding of an individuated system whose present experience is shaped by its own accumulated history. The emphasis therefore shifts from how information is predicted, broadcasted, or integrated at a given moment to how a system's unique history of interactions continuously shapes the character of its ongoing **experience**.

**Leveraging the opportunity: A general concept of consciousness**

Our dual-resolution framework allows us to assess whether current AI systems meet the criteria for consciousness. The first criterion, drawn from ITI, is whether an entity is individuated from an environment in a manner that secures informational autonomy and self-maintenance. The second criterion, extracted from the MTM, is whether the system exhibits hysteresis that supports epistemic resolution, namely a temporally extended process of updating that constitutes subjective experience. Importantly, these criteria should not be understood as sufficient conditions for consciousness. Informational individuation is proposed as a necessary ontological condition, while history-sensitive epistemic updating is proposed as a necessary dynamic condition. Their conjunction does not by itself settle whether a system is conscious; rather, it defines a constructive research program whose adequacy must be tested by whether such systems generate independently motivated operational markers of subjectivity, such as self-world differentiation, body-boundaries, and history-bound patterns of experience.

These criteria make it possible to evaluate current artificial systems without assuming in advance either that they are conscious or that consciousness is restricted to biological organisms. The point is not to ask whether present systems merely behave intelligently, but whether their organization supports the two conditions specified above: informational individuation and temporally extended epistemic updating. From this perspective, current AI systems occupy different positions. Standard LLMs under ordinary deployment conditions provide one limiting case, while environment-embedded reinforcement-learning agents provide another.

This contrast should be handled carefully. LLMs and RL agents are not parallel categories: LLMs are a class of models, whereas reinforcement learning is a training and control paradigm that can be combined with many architectures, including language models. The relevant comparison here is therefore between standard text-only LLM inference under ordinary deployment conditions and artificial agents embedded in environments in which their actions alter subsequent states.

Under ordinary deployment, standard LLMs exhibit local history-dependence: each generated token is conditioned on the preceding context, and generated outputs may become part of the subsequent input stream. Yet this dependence is typically confined to the current context window or to externally supplied memory systems. It does not by itself constitute an enduring informational individual. The system does not maintain itself as an autonomous unit over time, regulate its own boundary conditions, or preserve intrinsic viability variables whose maintenance constrains its future operation. From the MTM perspective, such systems may display local updating, but this updating is not bound to an individuated informational unit and therefore does not amount to epistemic resolution in the subjectivity-relevant sense.

Environment-embedded RL agents present a different case. Their actions can modify the state of the environment, and their behavior can be shaped by temporally extended feedback. This makes them closer to the class of systems relevant to ITI and MTM. However, standard episodic RL agents should not be treated as fully individuated in the ITI sense. Their boundaries, goals, reward functions, and persistence conditions are usually imposed by the designer, and policy change often occurs during training rather than as an ongoing self-maintaining trajectory during deployment. Thus, they display task-bounded agency without necessarily possessing informational autonomy. A stronger candidate would be a continual, embodied, homeostatic RL agent with persistent internal variables, intrinsic viability constraints, recurrent or memory-based updating, and the capacity to regulate its own relation to the environment over time.

Nevertheless, both types of systems could in principle be modified in ways that move them closer to the dual-resolution criteria. For LLMs, the relevant modification would not merely be a larger context window, memory module, or more coherent conversational persona. Rather, the system would require mechanisms of autonomous self-maintenance: persistent internal variables, regulated boundaries between internal and external states, intrinsic constraints whose preservation affects the system's continued operation, and forms of action through which the system can intervene on the conditions of its own persistence. Only under such conditions would sequential updating begin to acquire ontological grounding. For RL agents, the relevant modification would be the introduction of richer forms of hysteresis and self-maintenance. Their temporal dynamics would need to integrate context, goal trajectories, saliency weighting, and internally regulated viability variables, so that past interactions are not merely stored as parameters for reward maximization but actively shape how the system distinguishes itself from the environment. In this way, artificial systems could be designed not simply to optimize behavior, but to test whether informational individuation and moment-to-moment updating jointly generate operational markers of subjectivity.

At the same time, one might object that this line of reasoning risks circularity. If we begin by defining consciousness in terms of individuation and hysteresis, then it follows that any system engineered to satisfy those constraints will, by definition, count as conscious. This circularity would make the exercise uninteresting if left at the purely definitional level: it collapses into a tautology where "consciousness" simply names whatever meets the stipulated criteria. To avoid this pitfall, the ITI–MTM framework (or any other framework, for that matters) must be tested independently. We should not only retrofit machines to match our concepts. Instead, the goal would be testing whether meeting these criteria causes the emergence of core characteristics of consciousness.

**The opportunity AI advancement poses to the study of consciousness**

Rather than treating the question "are current AI systems conscious?" as decisive, AI advancement should be understood as creating a new methodological opportunity for consciousness science. Artificial systems allow researchers to construct, manipulate, and compare architectures in ways that are impossible or ethically constrained in biological organisms. They can therefore serve as testbeds not only for assessing whether particular systems are conscious, but also for examining whether our theories specify genuinely general conditions for consciousness.

One example of this approach is provided by Wilterson and Graziano (2021), who used AI to test the central proposal of Attention Schema Theory (AST). According to AST being aware of an item reflects the brain's representation of an internal model that describes how attention is currently allocated, such that the system attributes awareness not to the stimulus itself, but to an internal model that describes how attention is allocated towards this item. To test this proposal, the authors constructed an artificial agent tasked with controlling visuospatial attention and compared its performance with and without access to an explicit attention schema. They show that the agent performs significantly better at endogenous attentional control when it has access to a representation of its own attention, and that performance deteriorates when this representation is removed. Importantly, these results do not test a subjective characteristic of the AST model. Instead, it provides computational support for the core functional claim of AST by demonstrating that a model of attention confers a clear advantage for attentional control when implemented in an artificial system.

A similar approach is taken by Huang et al. (2022). They used AI to test the GNW by implementing its core mechanisms in a computational model. The model successfully reproduced classic attentional phenomena such as attentional blink and lag-1 sparing, supporting the idea that GNW identifies real functional mechanisms underlying conscious processing, rather than merely offering a post-hoc description of behavior. Here too, the artificial system does not reflect a subjective characteristic of GNW, but functions as a controlled platform for testing whether the mechanisms proposed by the theory can generate the cognitive phenomena observed in human participants.

Finally, Immertreu et al. (2025) used RL agents as a testbed for Damasio's theory of consciousness (Damasio & Damasio, 2022; Man & Damasio, 2019). Damasio's theory proposes that core consciousness depends on the integration of bodily or internal-state representations with representations of the external world. To test this, they trained RL agents in MiniHack environments and used probes to examine whether the agents' internal activations encoded their spatial position. Their results show that positional information could be decoded from the agents' neural states, especially in agents equipped with recurrent memory and restricted visual input. This suggests that artificial agents can acquire internal representations relevant to world modeling, and possibly to self-location, under appropriate architectural constraints. However, the findings should be interpreted cautiously: they demonstrate structural precursors to Damasio-style self–world modeling, not consciousness itself, and they do not yet show a full self-model or the hierarchical emergence of protoself, core consciousness, and extended consciousness.

Taken together, these studies demonstrate the productive role that artificial systems can play in consciousness science. They show that AI can be used not merely as an object of attribution, asking whether a given system is conscious, but as an experimental platform for implementing and testing theoretical claims about consciousness. Importantly, these studies differ in the extent to which they engage with subjective aspects of consciousness. The work of Wilterson and Graziano (2021) and Huang et al. (2022) primarily evaluates functional mechanisms proposed by AST and GNW, respectively, without directly addressing subjectivity itself. By contrast, the study by Immertreu et al. (2025) moves a step further by investigating whether artificial agents can develop internal self–world representations that may be considered precursors to subjective experience. In this sense, it begins to address not only the functional organization associated with consciousness but also structures that could underlie a first-person perspective.

Nevertheless, all three approaches remain tied to specific theoretical commitments. They test whether particular mechanisms or representational structures proposed by existing theories can emerge in artificial systems, thereby supporting or constraining those theories. What they do not provide is a more general account of the conditions under which subjectivity itself might arise across different kinds of systems. Thus, while these studies represent important advances (particularly the move toward self-related representations in the work of Immertreu et al.), they remain primarily tests of particular theories rather than tests of the general conditions under which subjectivity could emerge.

The dual-resolution framework extends this use of AI as a testbed by shifting the question from whether artificial systems instantiate isolated consciousness-related mechanisms to whether they satisfy the more general conditions under which consciousness could emerge. On this view, AI systems are valuable not only because they can implement components of existing theories, but because their architecture, memory, autonomy, embodiment, and temporal dynamics can be systematically manipulated. This makes it possible to test whether ontological individuation, as specified by ITI, and epistemic hysteresis, as specified by MTM, jointly give rise to operational markers of subjectivity, such as self–world differentiation, body-boundaries, and history-bound patterns of experience. The advantage of this proposal is therefore that it does not treat AI merely as a simulation of human consciousness or as a checklist for theory-derived indicators. Instead, it uses AI to examine whether the proposed general conditions for consciousness can generate conscious-like organization across radically different substrates.

This is where the dual-resolution framework makes a distinctive contribution. Previous AI-based tests have shown that artificial systems can instantiate and evaluate specific theoretical mechanisms. The present proposal asks a broader question: what must be added to an artificial system for it to become not merely an information processor, but an individuated system whose ongoing updates constitute a subjective trajectory? By defining consciousness through the conjunction of informational individuation and moment-to-moment epistemic updating, the ITI–MTM framework transforms AI from a passive target of consciousness attribution into an active experimental arena for testing the general conditions of subjectivity.

**AI as a testbed**

To explore whether conscious-like characteristics can emerge once both individuation and hysteresis are applied, we outline a simple experimental paradigm with modified RL agents. In this setup, the agents are endowed with mechanisms of informational autonomy, such as persistent self-governed memory and internal state variables that require ongoing maintenance, alongside hysteresis dynamics that modulate policy updates according to context, goals, and stimulus saliency. The central question is whether, under these conditions, the agents spontaneously acquire operational analogues of body boundaries, that is, the capacity to distinguish between self-generated and externally imposed perturbations and to allocate resources preferentially to the protection of their own integrity. For example, agents will be placed in a simulated environment containing manipulable objects, hazards, and opportunities for self-repair. Crucially, some perturbations of the agent's "body" are self-caused, while others are externally imposed by the environment. If the agent anticipates the sensory consequences of its own actions, treats them as less surprising, and responds more vigorously to external disruptions, this will indicate the emergence of a functional boundary between self and world. Over time, tools can be introduced to test whether the agent flexibly incorporates new extensions into its body representation, ceasing to treat tool-generated effects as external, once they are under its control.

What makes this approach non-trivial is that it yields empirical consequences. If individuated systems with enriched hysteresis display novel behaviors, such as spontaneous emergence of body boundaries, then the framework gains explanatory traction. Conversely, if no such behaviors appear despite engineering efforts, this provides evidence against the sufficiency of our criteria for consciousness. The framework thus avoids circularity by treating AI systems as experimental probes. Moreover, it corresponds to Crick and Koch's (1990) notion of reciprocal relations between empirical results and theory, as any result from these probes would feedback into the theory itself.

Finally, this experimental strategy highlights a broader methodological shift. Instead of trying to declare which entities "are conscious," we treat consciousness research as a program of constructive exploration. By iteratively "playing" with architectures, constraints, and environments, we can generate falsifiable predictions and refine our conceptual tools. In this sense, AI serves as testbeds for the science of consciousness, not only for examining its capacities, but also the robustness and universality of our theoretical frameworks for consciousness.

**Conclusion**

The encounter with AI provides a unique opportunity for the science of consciousness. Unlike earlier comparative cases such as animal consciousness, AI compels us to examine consciousness in a substrate-independent manner. The debate is not merely whether artificial systems can be conscious, but whether our theories are sufficiently general to accommodate forms of subjectivity that may differ radically from our own.

In this paper, we have proposed a dual-resolution framework that synthesizes the Information Theory of Individuality (ITI) and the Moment-to-Moment (MTM) theory. ITI secures the ontological conditions of individuality, defining the informational autonomy necessary for any conscious entity, while MTM secures the epistemic conditions by explaining how subjective experience arises through continuous updating and integration of information. Together, these perspectives suggest that consciousness is best understood as the epistemic unfolding of individuated life, regardless of whether the substrate is biological or artificial.

This framework has two major implications. First, it reframes the challenge of AI to the science of consciousness as a diagnostic test of its scope and validity. A valuable theory of consciousness should accommodate and explain the general conditions in which it could emerge and avoid an anthropocentric view. Second, it points toward a program of research where AI is treated as an experimental companion: a context in which new models of information processing, self-maintenance, and temporal updating can be directly implemented, tested, and compared against biological consciousness.

Looking ahead, the development of consciousness science will require abandoning anthropocentrism and embracing models that are explicit about their substrate independence. The ITI–MTM dual-resolution perspective represents one such attempt. We do not claim it is final or exhaustive; rather, it demonstrates how combining ontological and epistemic criteria can open a path toward theories capable of generalizing across various entities. By engaging with AI as both a challenge and an opportunity, consciousness science can evolve into a more rigorous, universal discipline - one that does not merely describe the human case but seeks to explain consciousness *wherever it may arise*.

Acknowledgements: This work was supported by ISF grant #2680/24 for authors DB and MS